\documentclass{amsart}

\usepackage{amssymb}

\mathchardef\ordinarycolon\mathcode`\:     % this is for a correctly aligned :=
\mathcode`\:=\string"8000
\def\vcentcolon{\mathrel{\mathop\ordinarycolon}} \begingroup
\catcode`\:=\active \lowercase{\endgroup \let :\vcentcolon }

\theoremstyle{definition}
\newtheorem{example}{Example}

\newtheorem{definition}{Definition}
\newtheorem{algorithm}{Algorithm}

\theoremstyle{plain}

\newtheorem{theorem}{Theorem}

\newcommand{\V}{{\textbf V}}  
\newcommand{\I}{{\textbf I}}  
\newcommand{\U}{{\mathsf U}}  
\newcommand{\Z}{{\mathbb Z}}  
\newcommand{\C}{{\mathbb C}} 
\newcommand{\Q}{{\mathbb Q}} 
\renewcommand{\P}{{\mathbb P}} 
\newcommand{\F}{{\mathbb F}} 
\newcommand{\A}{{\mathbb A}} 
\newcommand{\ket}[1]{|#1\rangle} 
\newcommand{\braket}[2]{\langle#1|#2\rangle} 
\newcommand{\e}{\mathrm{e}}
\renewcommand{\i}{\mathrm{i}}
\renewcommand{\d}{\mathrm{d}}
\newcommand{\poly}{\mathrm{poly}}
\newcommand{\Tr}{\mathrm{Tr}}
\renewcommand{\root}[2]{\omega_{#1}^{#2}}
\newcommand{\smfrac}[2]{\mbox{$\frac{#1}{#2}$}}

\newcommand{\Zeta}{\mathrm{Z}}
\newcommand{\years}[1]{\oldstylenums{#1}}
\newcommand{\sharpP}{\mathsf{\#{}P}}

\begin{document}

\title[Quantum Computing and Zeroes of Zeta Functions]%
{Quantum Computing and Zeroes of Zeta Functions}
\author{Wim van Dam}
\address{~\newline \indent Wim van Dam 
\newline\indent Massachusetts Institute of Technology
\newline\indent Center for Theoretical Physics
\newline\indent {77} Massachusetts Avenue
\newline\indent Cambridge, MA {02139-4307}, USA}
\thanks{%
Report no.~MIT-CTP 3494. 
This work is supported in part by funds provided by the 
U.S.\ Department of Energy  and cooperative research 
agreement DF-FC02-94ER40818, by a CMI postdoctoral fellowship
and by an earlier HP/MSRI fellowship.
}  
\email{vandam@mit.edu}


\begin{abstract}
A possible connection between quantum computing and Zeta functions of 
finite field equations is described. 
Inspired by the \emph{spectral approach} to the Riemann conjecture, 
the assumption is that the zeroes of such Zeta functions correspond to the
eigenvalues of finite dimensional unitary operators of natural quantum
mechanical systems.  The notion of universal, efficient quantum computation
is used to model the desired quantum systems. 
 
Using eigenvalue estimation, such quantum circuits would be able to
approximately count the number of solutions of finite
field equations with an accuracy that does not appear to be feasible
with a classical computer. For certain equations (Fermat hypersurfaces) it is 
show that one can indeed model their Zeta functions with efficient quantum
algorithms, which gives some evidence in favor of the proposal
of this article.
\end{abstract}

\maketitle

\newpage
\tableofcontents 
\newpage

\section{Introduction}
The \emph{spectral approach} to the Riemann conjecture tries
to interpret the zeroes of the Riemann zeta function as 
the eigenvalues of a physical operator, like a Hamiltonian
in quantum mechanics.  This idea goes back, apparently,
to Hilbert and P{\'o}lya, although neither have published 
about this.  Evidence supporting the spectral interpretation
comes from the observed spacing of the zeroes of Riemann's zeta 
function, which has a striking resemblance with the `eigenvalue 
repulsion' of chaotic quantum mechanical systems\cite{Odlyzko}.
An explicit construction of an operator whose spectrum
coincides with the zeroes of $\zeta$ would show that these roots 
indeed lie on a line, thereby proving Riemann's conjecture.
See \cite{BK} for an attempt to discover a Hamiltonian 
with such properties.

In this article we consider a different kind of Zeta functions:
those that correspond to equations over finite fields.  
Such Zeta functions have a finite number of roots, 
and the corresponding `Riemann hypothesis' of Andr\'e 
Weil states that these roots lie on a circle in the 
complex plane.  Unlike the case of the Riemann zeta 
function, this hypothesis has been proven. 
Here we propose to interpret the roots of such 
Zeta functions as eigenvalues of unitary 
transformations
of finite dimensional quantum systems.  More specifically
we want these quantum systems to be natural, which we will
define to mean that they correspond to  efficient 
quantum algorithms.
By the properties of these Zeta functions, such algorithms 
would enable us to estimate the number of solutions of the 
equations with an accuracy that does not seem to be 
possible classically.

Using earlier results on the quantum estimation of 
Gauss sums, we show that for certain diagonal equations, 
also known as Fermat hypersurfaces'',  
we can indeed construct a corresponding quantum circuit.
It is an open problem how to extend this result
to general hypersurfaces and, ultimately, varieties.

\subsubsection*{Notation}
Throughout the article we use the following conventions.  The number
of variables $X$ in a polynomial $f$ is $n$, such that in the affine
$\A^n$ setting $f\in\F[X_1,\dots,X_n]$ and in the projective case
$\P^n$ we have the homogeneous polynomial $f\in\F[X_0,\dots,X_n]$ of
degree $m$.  The finite field $\F$ has $q=p^r$ elements and its
extensions $\F_{q^s}$ are indexed by $s$.  The dimension of a variety
$V$ is denoted by $d$, such that for hypersurfaces defined by $f$ we
have $d=n-1$.  If an ideal is defined by several polynomials $f$, then
the number of polynomials will be $t$.  The variable of a zeta
function $\zeta$ is $z$, while those of Zeta functions $\Zeta$ will be
$T$.

\section{Quantum Computing}
In this Section we briefly describe the basic ingredients 
of the theory of quantum computation.
See \cite{Kitaev} or \cite{NC} for a thorough introduction 
to this field.
\subsection{Quantum States and Quantum Transformations}
A quantum state $\psi$ of $n$ quantum bits (qubits) is described 
by a $2^n$ dimensional complex valued vector, which represents
a superposition over all possible $n$ bit strings $\{0,1\}^n$.
In the `bra-ket' notation this is expressed as
\begin{eqnarray*}
\ket{\psi} & = & \sum_{x\in\{0,1\}^n}{\alpha_x\ket{x}},
\end{eqnarray*}
with $\alpha_x\in\C$ and the normalization restriction 
$\sum_x{|\alpha_x|^2}=1$.  When observing this state $\psi$
in the computational basis $\{0,1\}^n$, the probability
of observing a specific string $x\in\{0,1\}^n$ is $|\alpha_x|^2$
(hence the normalization restriction).
In general, the probability of observing the state 
${\phi}$ when performing a measurement on $\psi$
is the `inner product squared' $|\braket{\phi}{\psi}|^2$.

A coherent quantum mechanical transformation $T$ can always be 
described by a linear transformation that preserves the 
normalization restriction of the state vectors.  Thus the 
transformation $T$ of an $n$ qubit system can be represented
as a $2^n\times 2^n$ dimensional unitary matrix: 
$T\in\U(2^n)$. Any such transformation is reversible as $T^\dagger\cdot T=I$.
 A $k$-qubit quantum gate $G$ is a unitary transformation of $k$ qubits, 
such that $G\in \U(2^k)$.  

\subsection{Efficient Quantum Algorithms}
A quantum circuit $C\in\U(2^n)$ on $n$ qubits can be defined as sequence 
of $2$ qubit gates (note that this includes single qubit
gates and the two-qubit \textsc{swap} operation for the 
wiring of the circuit).  The outcome of a circuit on 
a given input string $x\in\{0,1\}^n$ is the probability 
distribution of the output state $C\ket{x}$ over the computational 
basis states $\{0,1\}^n$. The depth of such a circuit
equals its time complexity and we consider a family of circuits
efficient if its depth as a function of $n$ is upper bounded by 
$O(\poly(\log n))$.

Almost all quantum algorithms known to date rely on the properties
and the efficiency of the quantum Fourier transform.
\begin{definition}[{Efficient Quantum Fourier Transformation}]
For a ring $\Z/m\Z$, the quantum Fourier transformation $F$
is defined  as the unitary mapping
\begin{eqnarray*}
F:\ket{x} & \longmapsto & \frac{1}{\sqrt{m}}\sum_{y=0}^{m-1}{\root{m}{xy}\ket{y}},
\end{eqnarray*}
with $\root{m}{} := \e^{2\pi\i/m}$.

For a finite field $\F_{p^r}$ 
the corresponding quantum Fourier transformation is defined by 
\begin{eqnarray*}
F:\ket{x} & \longmapsto &
\frac{1}{\sqrt{p^r}}\sum_{y\in\F_{p^r}}{\root{p}{\Tr(xy)}\ket{y}}
\end{eqnarray*}
for every $x\in \F_{p^r}$, where we used the trace function 
$\Tr:\F_{p^r}\rightarrow \F_p$ defined by 
$\Tr(x) := x+x^{p}+\cdots+x^{p^{r-1}}$. 
\end{definition}
It is well-known that both quantum Fourier transforms can be 
implemented efficiently on a quantum computer 
(with circuit depth $\poly(\log n)$ and 
$\poly(\log p^r)$ respectively \cite{vDS1,vDS2,revisited}).
This fact is an important ingredient of the efficient quantum algorithms 
for factoring and the discrete logarithm problem\cite{Shor}.

\subsection{Universality of Efficient Quantum Computation}\label{sec:uni}
An important aspect of the theory of efficient quantum computation is
that it is independent of the computational model one uses.  Just as
with the classical theory of polynomial-time computation, it does not
matter if one expresses it in terms of Turing machines, circuits, or
in any other reasonable model.  This universality is proven by, for
example, the fact that a quantum Turing machine can efficiently
simulate a quantum circuit and vice-versa.  Furthermore we have the
modern Church-Turing thesis, which says that the physical resources
that are required to solve a computational problem are properly
quantified by the space/time complexity of the problem in the abstract
quantum Turing machine model.  Again, see \cite{Kitaev,NC} for more
details.

\section{Zeta Functions of Polynomial Equations in Finite Fields}
\subsection{Projective Spaces, Hypersurfaces and Varieties}
Consider a homogeneous polynomial $f\in\F_q[X_0,\dots,X_n]$
and its solutions obeying $f(x)=0$.  The preferred way of analyzing 
such a set of solutions is with the use of the projective 
space $\P^n$.
\begin{definition}[{Projective Spaces}]
The \emph{projective space} $\P^{n}(\F)$ is the set of 
distinct rays
$x = \{(\lambda x_0,\lambda x_1,\dots,\lambda x_n):\lambda\in\F^\times\}$,
excluding the zero-point $(0,\dots,0)$. 
If $|\F|$ is finite, each ray contains $|\F|-1$ points from the
affine space $\A^{n+1}(\F)$.  This affine space itself has
$|\F|^{n+1}-1$ non-zero points, hence $\P^{n}(\F)$ has
$(|\F|^{n+1}-1)/(|\F|-1) = |\F|^n + |\F|^{n-1} + \cdots + 1$ elements.
\end{definition}
In algebraic geometry, projective spaces are often preferred over
affine ones because $\P^{n}$ includes the `points at infinity' such
that it is a closed space.

By the homogeneity of the polynomial $f$ it follows that if
$f(x_0,\dots,x_n)=0$ then also $f(\lambda x_0,\dots,\lambda x_{n})=0$
for all $\lambda$.  We can therefore group the nontrivial roots of $f$
as rays in $\P^{n}$ and hence speak of the set of solutions of
$f(x)=0$ with $x\in\P^{n}(\F)$.

The algebraic closure of $\F_q$ is denoted by $\bar{\F}_q$, which is
the infinite field that contains all extensions $\F_{q^s}$ of
$\F_q$. For $f\in\F_q[X_0,\dots,X_{n}]$ we want to investigate the set
of solutions $H=\{x:x\in\P^{n}(\bar\F_q) \text{ and }f(x)=0\}$, which
for obvious reasons we will call an $(n-1)$-dimensional hypersurface.

We can generalize the notion of a hypersurface by considering the
joint solutions to several polynomial equations $\{f_i(x)=0\}_i$.
Such an algebraic set is called a \emph{variety} in the context of
algebraic geometry and is thus defined by $V=\{x:x\in\P^{n}(\bar\F_q),
f_1(x)=0,\dots,f_t(x)=0\}$.  A variety has a dimension, varieties of
dimension $1$ are called \emph{curves,} and curves that are defined by
a single polynomial $f$ are \emph{planar curves.}

\subsection{Zeta Function of a Variety}
Given a variety $V$ of a set of polynomials $\{f_1,\dots,f_t\}$, we
can look at the solutions that $V$ has in the finite extensions
$\F_{q^s}$ of the finite fields $\F_q$.  That is, we consider the
sizes of the intersections $|V\cap \P^{n}(\F_{q^s})|$ for
$s=1,2,\dots$.  This brings us to the following important definition.
\begin{definition}[{Zeta Function of a Variety}]
Let $V$ be a projective variety defined over $\P^{n}(\bar\F_q)$ and
let $N_s := |V\cap \P^{n}(\F_{q^s})|$ for $s=\Z^+$.  The \emph{Zeta
function} of the variety $V$ is the power series $\Zeta_V(T)\in
\Q[\![T]\!]$ defined by
\begin{eqnarray*}
\Zeta_V(T) &:=&  
\exp\left(\sum_{s=1}^{\infty}{\frac{N_s T^s}{s}}\right).
\end{eqnarray*}  
\end{definition}
Why it is a good idea to define a function like this and 
why one can call it a Zeta function is not entirely trivial.  
The reader is referred to the appendices in this article and 
the many books and articles that deal with this topic, in 
order of increasing difficulty, 
\cite{IR,Rosen,Lorenzini,Thomas,Harts}.
For now, we will make do with a few small examples.

\subsection{Three Examples of Zeta Functions of Hypersurfaces}
\begin{example}[{Zeta Function of a Straight Line}]\label{ex:line}
Consider the line 
$L\subset \P^2(\bar\F_q)$ defined by $f(X_0,X_1,X_2) = X_1+X_2 = 0$. 
The solutions to this equation are the rays
$\{(\lambda,0,0):\lambda\in\F^\times\}$ and 
$\{(\lambda z,\lambda,-\lambda):\lambda\in\F^\times\}$ for every 
$z\in \F$. Hence $L$ has $1+|\F|$ solutions, and thus $N_s=1+q^s$.
For the Zeta function $\Zeta_L$ this means 
\begin{eqnarray*}
\Zeta_L(T) & =& \exp\left(\sum_{s=1}^{\infty}{\frac{(1+q^s) T^s}{s}}\right)\\
& = & \exp\left(\sum_{s=1}^{\infty}{\frac{T^s}{s}}\right)
\exp\left(\sum_{s=1}^{\infty}{\frac{(qT)^s}{s}}\right) \\
& = & \frac{1}{(1-T)(1-qT)}.
\end{eqnarray*}
Clearly every straight line with $N_s = 1+q^s$ will have this Zeta function.
\end{example}
\begin{example}[{Zeta Function of a 2 Dimensional Hypersurface, Section~11.1 in \cite{IR}}]
Take the hypersurface $H$ defined by 
$f(X_0,\dots,X_3)=-X_0^2+X_1^2+X_2^2+X_3^2 = 0$
over $\P^3(\bar\F_7)$. One can show that for this
specific base field $\F_7$ we have $N_s = 1+7^s +(-7)^s + 49^s$, 
and thus 
\begin{eqnarray*}
\Zeta_H(T) & =& \exp\left(\sum_{s=1}^{\infty}{\frac{(1+7^s + (-7)^s+ 49^s) T^s}{s}}\right)\\
& = & \frac{1}{(1-T)(1-49T^2)(1-49T)}.
\end{eqnarray*}
\end{example}
\begin{example}[Zeta Function of a Quartic Diagonal Equation]\label{ex:curve}
Let $f(X_0,X_1,X_2) = X_0^4 + X_1^4 + X_2^4$ define a curve $C$ in 
$\P^2(\bar\F_5)$.  Its Zeta function equals
\begin{eqnarray*}
\Zeta_C(T) & = & \frac{(1-2T+5T^2)^3}{(1-T)(1-qT)}.
\end{eqnarray*}
Note that the roots of the denominator are $T=\frac{1}{5}\pm \frac{2}{5}\i$.
\end{example}
The kind of regularity of the above examples holds 
for all varieties and is expressed in the important 
\emph{Weil conjectures} \cite{Dieu,Rosen,Weil,Zeta},
which have been proven during the period \years{1949}--\years{1973}.
As a result we now know that $\Zeta_V(T)$ is always a rational 
function; but the most important part of the Weil 
conjectures concerns the roots and poles of the Zeta functions.  

\subsection{Weil's Riemann Hypothesis}
For concreteness, we first describe part of 
Weil's conjecture regarding the Zeta function of curves. 
\begin{theorem}[{Weil's Riemann Hypothesis for Curves \cite{Weil}}]
Let $C$ be a complete, nonsingular curve over $\F_q$.  
The corresponding Zeta function $\Zeta_C$ is a rational function 
$\in \Z(T)$ that can  be decomposed as
\begin{eqnarray*}
\Zeta_C(T) & = & \frac{P(T)}{(1-T)(1-qT)} \\
& = & 
\frac{\prod_{j=1}^{2g}(1-\alpha_{j}T)}{(1-T)(1-qT)},
\end{eqnarray*} 
where $g$ is the \emph{genus} of $C$.
The polynomial $P(T)$ has integer coefficients: $P(T)\in\Z[T]$,
and the roots of $P(T)$ can be grouped in pairs $(\alpha_j,\alpha_{j+g})$ 
such that $\alpha_j = \bar{\alpha}_{j+g}$.
Most importantly, the magnitudes of the roots of $P(T)$ obey 
$|\alpha_{j}|=\sqrt{q}$ for all $j$.
\end{theorem}
The genus of a curve reflects how different $C$ is from a straight 
line (which has genus $0$).  This quantity has a close connection
with the \emph{geometric genus} of the curve $f=0$ when viewed
in the projective space $\P^2(\C)$.
Typically, the genus of a curve defined by $f$ is approximately 
$\deg(f)^2$.

Example~\ref{ex:curve} is a typical instances of this theorem: The
Zeta function of the curve $X_0^4 + X_1^4 + X_2^4=0$ in
$\P^2(\bar\F_5)$ has $6$ roots, all with norm $1/\sqrt{5}$, and when
we view the equation in the complex projective space in $\P^2(\C)$ we
see that it has geometric genus $3$, which corresponds with the fact
that its Zeta functions has $2g=6$ roots.

The most general version of Weil's Riemann hypothesis applies to all
`proper' varieties.
\begin{theorem}[{Weil's Riemann Hypothesis for Varieties, \cite{Weil}}]
\label{thm:weil}
Let $V$ be a complete, nonsingular algebraic variety of dimension $d$
over $\F_q$.  The corresponding Zeta function $\Zeta_V$ is a rational
function that can be decomposed as
\begin{eqnarray*}
\Zeta_V(T) & = & \frac{P_1(T)\cdots P_{2d-1}(T)}{P_0(T)P_2(T)\cdots P_{2d}(T)} \\
& = & 
\frac{\prod_{j=1}^{B_1}(1-\alpha_{1,j}T)\cdots\prod_{j=1}^{B_{2d-1}}(1-\alpha_{2d-1,j}T)}
{\prod_{j=1}^{B_0}{(1-\alpha_{0,j}T)}\prod_{j=1}^{B_2}{(1-\alpha_{2,j}T)}%
\cdots\prod_{j=1}^{B_{2d}}{(1-\alpha_{2d,j}T)}},
\end{eqnarray*} 
where $B_i$ are the \emph{Betti numbers} of $V$. 
All the polynomials $P_i(T)$ are elements of $\Z[T]$, the roots of the 
$P_i$ polynomials can be paired such that $\alpha_{i,j}=\bar{\alpha}_{i,B_i/2+j}$.
It always holds that $P_0(T)=1-T$ and $P_{2d}(T)=1-q^d T$, and 
in general the magnitudes of these roots obey 
$|\alpha_{i,j}|={q}^{i/2}$ for all $i,j$.
\end{theorem}
As stated before, this theorem has been proven; see 
\cite{Dieu,Harts,Lorenzini,Rosen,Thomas,Zeta}.

For hypersurfaces, which are defined by a single polynomial, 
the Zeta functions is particularly simple.
Let $f\in\F[X_0,\dots,X_n]$ be a homogeneous polynomial 
 that defines a proper hypersurface $H$, 
then its Zeta function has the form
\begin{eqnarray}\label{eq:zetahyper}
\Zeta_H(T) & = & \frac{P(T)^{(-1)^{n}}}{(1-T)(1-qT)\cdots (1-q^{n-1}T)},
\end{eqnarray}
where each of the $[(\deg(f)-1)^{n+1}-(-1)^{n}(\deg(f)-1)]/\deg(f) \approx
\deg(f)^n$ roots of $P(T)$ have magnitude $q^{-(n-1)/2}$.

\section{Zeroes of Zeta Functions as Eigenvalues of Quantum Circuits}
We want to interpret the roots of Zeta functions of the previous
section as the spectrum of a quantum mechanical process.
This wish is a natural extension of the spectral approach
to the Riemann conjecture with the added advantage that we already 
have more evidence in favor of it, especially in the case of curves.

First of all, we know that Weil's Riemann hypothesis is true, hence we
 know that the roots of the Zeta functions all lie on a circle in the
 complex plane.  Second, it has also been proven by Katz and Sarnak
 \cite{KZ1,KZ2} that for curves the distribution of the zeroes of the
 Zeta functions obeys the kind `eigenvalue repulsion' that one also
 sees in random quantum mechanical systems.  Specifically, they showed
 that for `generic' curves $C$ in the double limit $g\rightarrow
 \infty$ and $q\rightarrow \infty$ the distribution of the zeroes of
 $\Zeta_C$ goes to the eigenvalue distribution of the circular unitary 
(symplectic) ensemble. (See also \cite{KLR}.)

Of course, given a curve $C$, one can always define a diagonal
unitary matrix with as its diagonal entries the normalized
roots of $\Zeta_C$ and declare this matrix to describe a quantum
mechanical system.  Clearly this is not a satisfactory answer
because this definition does not give a natural description 
of a physical system.  Hence, before we can proceed, we have to 
find a criterion for what does constitute a natural physical 
system. 
 
Here we propose to use the class of efficient quantum algorithms to
determine if a sequence of unitary operations is considered natural or
not.  Formally, a sequence $U_1,U_2,\dots$ of unitary matrices with
$U_i\in\U(N_i)$ is natural if and only if there is a quantum algorithm
that for given $i$ efficiently implements the transformation $U_i$ in
time $O(\poly(\log N_i))$.  More concretely, for a variety $V$ we want
to describe an efficient quantum circuit whose eigenvalues correspond
to the phases of the zeroes of the Zeta function $\Zeta_V$.  The
earlier described universality results in Section~\ref{sec:uni} show
that this criterion is independent of the specific quantum
computational model that we use, and the quantum Church-Turing thesis
suggests that it captures exactly those quantum mechanical systems
that can occur in Nature without some kind of exponential overhead.

The next section shows how for so called \emph{Fermat hypersurfaces}
$H$, which are defined by diagonal equations $c_0 X^m_0 + c_1X_1^m +
\cdots + c_{d+1} X_{d+1}^m = 0$, we can indeed construct an efficient
quantum algorithm that has the normalized roots of $\Zeta_H$ as its
eigenvalues.

\section{Quantum Algorithm for Zeta Functions of Some Hypersurfaces}
For some hypersurfaces $H$ we can express the roots of the
corresponding Zeta functions $\Zeta_H$ in terms of products of Gauss
sums.\cite{IR,Koblitz} Before we give the quantum algorithm whose
eigenvalues are these roots, we will repeat some known results on the
quantum computation of Gauss sums.

\subsection{Multiplicative Characters and Quantum Computing Gauss Sums}
A \emph{multiplicative character} of a finite field is a function
$\chi:\F_{q}\rightarrow \C$ such that $\chi(xy)=\chi(x)\chi(y)$ for
all $x,y\in\F_{q}$.  Let $g$ be a primitive element of $\F_{q}$, i.e.\
the multiplicative group $\langle g \rangle$ generated by $g$ equals
$\F^\times_{q}:=\F_{q}\!\!\setminus\!\!\{0\}$, and let $\root{q-1}{}
:= \e^{2\pi\i/(q-1)}$.  For each $\alpha \in\{0,1,\dots,q-2\}$, the
function $\chi(g^j) := \root{q-1}{\alpha j}$ (complemented with
$\chi(0) := 0$) is a multiplicative character.  Conversely, every
multiplicative character can be written as such a function.  The
\emph{discrete logarithm} with respect to $g$ is defined for every
$x=g^j\in\F_{q}^\times$ by $\log_g(x):=j\bmod{(q-1)}$.  Hence, every
multiplicative character can be expressed as $\chi(x) :=
\root{q-1}{\alpha\log_g(x)}$ for $x\neq 0$ and $\chi(0):=0$.

Fix a generator $g$ and define the primitive multiplicative character
$\chi$ by $\chi(x) := \root{q-1}{\log_g(x)}$.  For every $\alpha\in\Z$
we have the $\alpha$-th power of $\chi$ according to $\chi^\alpha(x)
:= (\chi(x))^\alpha$.  Thus the \emph{trivial multiplicative
character} is denoted by $\chi^{0}$ and is defined by $\chi^{0}(0)=0$
and $\chi^{0}(x)=1$ for all $x\neq 0$.  Using the equality
$\chi^\alpha\cdot\chi^\beta = \chi^{\alpha+\beta}$ we see that the set
of characters $\{\chi^{\alpha}:\alpha\in\{0,\dots,q-2\}\}$ with
pointwise multiplication defines a group isomorphic to the additive
group $\Z/(q-1)\Z$.  The inverse of $\chi$ obeys $\chi^{(-1)}(x) =
\overline{\chi(x)}$ for all $x$, where $\bar{z}$ denotes the complex
conjugate of $z$.

\begin{definition}[{Gau{ss} sums over Finite Fields}]
For a finite field $\F_{q}$ with $q=p^r$ and a multiplicative
character $\chi$, we define the complex valued \emph{Gau{ss} sum} $g$
by
\begin{eqnarray*}
g(\chi) & := &
\sum_{x\in\F_{q}}{\chi(x)\root{p}{\Tr(x)}},
\end{eqnarray*}
where $\Tr$ is the standard trace function.
Obviously, $g(\chi^0) = -1$ and for nontrivial characters $\chi$
we have that the norm of the Gau{ss} sum obeys $|g(\chi)| = \sqrt{q}$. 
\end{definition}

To be a able to quantum compute we use the following states \cite{vD}, 
which were also previously described by Watrous in \cite{Watrous}.
\begin{definition}[{Chi States}]
Given a finite field $\F_{q}$ and a generator $g\in \F_{q}^\times$ 
we define the \emph{chi states} for every $\alpha\in\Z$ by 
\begin{eqnarray*}
\ket{\chi^\alpha} & := & \frac{1}{\sqrt{q-1}}\sum_{x\in\F_q}{\chi^\alpha(x)\ket{x}},
\end{eqnarray*} 
where $\chi^\alpha$ refers to the multiplicative character defined 
by $g$ in $\chi^\alpha(g^j)=\root{q-1}{\alpha j}$.
Note that $\ket{\chi^0}$ is the uniform superposition
of the elements of $\F_{q}^\times$.
Chi states can be produced in time $O(\poly(\log q))$ on a 
quantum computer.\cite{vD}
\end{definition}

The assumption that multiplication and division in $\F_{q}$ can be
done efficiently implies that, using repeated powering $x\mapsto
x^2\mapsto x^4\cdots$, we can efficiently calculate any power $x^j$ in
$\F_q$ for $-q < j < q$.  This is a useful operation in combination
with $\chi$ states.  It is straightforward to check that if we apply
the reversible $\ket{x,y}\mapsto \ket{x,y/x^\alpha}$ mapping (for
$x,y\in\F_q^\times$) to the superposition
$\ket{\chi^\beta}\ket{\chi^\gamma}$, then we obtain
$\ket{\chi^{\beta+\alpha\gamma}}\ket{\chi^\gamma}$.  Hence, under the
assumption that it is easy to create the uniform superposition
$\ket{\chi^0}$, we can efficiently create arbitrary
$\ket{\chi^\alpha}$ states from an initial state $\ket{\chi}$.
Chi states are especially helpful if we want to induce a phase change
$\e^{\i \theta}$ that is determined by a Gauss sum $g(\chi) = \e^{\i
\theta}\sqrt{q}$ of a nontrivial character $\chi$.  For, if we create
the state $\ket{\chi}$ and perform the Fourier transform over $\F_q$
to it, we implement the evolution $\ket{\chi} \mapsto
g(\chi)/{\sqrt{q}}\cdot\ket{\chi^{-1}}$.  See \cite{vDS1,vDS2} and
references therein for more detailed information about these topics.

\subsection{Quantum Algorithm for Zeta Functions of Fermat Hypersurfaces}
With the results of the previous subsection we have the following theorem.
\begin{theorem}\label{thm:qa}
Let $f$ be a homogeneous polynomial in $\F_q[X_0,\dots,X_{n}]$: 
\begin{eqnarray*}
f(X_0,\dots,X_{n}) & = & c_0 X_0^m + c_1X_1^m + \cdots + c_{n} X_{n}^m,
\end{eqnarray*}
with $q = 1\bmod{m}$, and let $H$ be the corresponding $(n-1)$-dimensional 
projective hypersurface $H := \{x: x\in\P^{n}(\bar{\F}_q)\text{ and } f(x)=0\}$. 
For every such $H$ there exists an efficient quantum circuit whose
eigenvalues are the roots of the Zeta function $\Zeta_H$.
\end{theorem}
\begin{proof}
In Chapter~11 of \cite{IR} it is explained how the the 
corresponding Zeta function equals  
\begin{eqnarray*}
\Zeta_H(T) & = & \frac{P(T)^{(-1)^{n}}}{(1-T)(1-qT)\cdots (1-q^{n-1}T)},
\end{eqnarray*}
where  $P$ (and hence $\Zeta_H$) has $(m-1)[(m-1)^{n}+(-1)^{n-1}]/m 
\approx m^{n}$ non-trivial zeroes, all on the circle $|T|=q^{-(n-1)/2}$.  

Define $\tilde\chi$ to be the character $\tilde\chi(g^j) := \root{q-1}{j(q-1)/m}$,
with $g$ a generator of $\F_q^\times$, such that $\tilde\chi^m=\chi^0$.
Then, we also know that the roots $1/\alpha_j$ of $P = \prod_j{(1-\alpha_jT)}$ are 
described by 
\begin{eqnarray}\label{eq:fermatroots}
\frac{1}{\alpha_j} & = & \frac{(-1)^{n-1}}{q^{n}}\cdot
\frac{g(\chi_0) \cdots g(\chi_{n})}{\chi_0(c_0) \cdots \chi_{n}(c_{n})},
\end{eqnarray}
where the multiplicative characters are defined by $\chi_i := \tilde\chi^{b_i}$,
with $b_i \in \{1,\dots,m-1\}$ for all $i$ and $\sum_i{b_i}=0\bmod{m}$.

We rewrite the roots as $1/\alpha_j = \e^{\i\theta_j}/\sqrt{q^{n-1}}$,
such that we can focus on the unknown angles $\theta_j \in [0,2\pi)$.   
Using the quantum algorithms for Gauss sum estimation and multiplicative
character phase changing described in \cite{vDS1,vDS2}, we can 
implement the evolution $\ket{b_0,\dots,b_{n}}\mapsto
\e^{\i\theta_j} \ket{b_0,\dots,b_{n}}$ for every $j$ and its 
corresponding sequence $(b_0,\dots,b_{n})$.
Hence this evolution has $\ket{b_0,\dots,b_{n}}$
as its eigenstates and the normalized roots $\sqrt{q^{n-1}}/\alpha_j = \e^{\i\theta_j}$ 
as the eigenvalues.
\end{proof}

In order to make this theorem more explicit, we describe the 
quantum algorithm (note that the size of the input $f$ is 
$\approx n\log q$ bits). 
\begin{algorithm}\label{alg:qa}
Given the polynomial $f$ of Theorem~\ref{thm:qa}, implement the 
following quantum evolution on a state $\ket{b_0,\dots,b_{n}}$
that obeys $b_i \neq 0$ for all $i$ and $\sum_i{b_i}=0\bmod{m}$.
\begin{enumerate}
\item Attach the chi states $\ket{\chi^0,\tilde{\chi}}$,  
giving $\ket{b_0,\dots,b_{n}}\otimes\ket{\chi^0,\tilde\chi}$.
\item For every $0\leq i \leq n$, perform the following steps:
\begin{enumerate}
\item Apply $\ket{x,y}\mapsto \ket{x,y/x^{b_i}}$ to $\ket{\chi^0,\tilde{\chi}}$ 
such that we get $\ket{\tilde\chi^{b_i},\tilde{\chi}}$.
\item Apply a Fourier transformation over $\F_q$ to the $\tilde\chi^{b_i}$ register, 
yielding the state $g(\tilde\chi^{b_i})/\sqrt{q}\cdot\ket{\tilde{\chi}^{-b_i},\tilde\chi}$.
\item Again perform $\ket{x,y}\mapsto \ket{x,y/x^{b_i}}$ to the
two chi registers such that the net effect of the subroutine is
$\ket{\chi^0,\tilde\chi} \mapsto 
g(\tilde\chi^{b_i})/\sqrt{q}\cdot\ket{{\chi}^0,\tilde\chi}$.
\end{enumerate}
We now have the state
$g(\chi_0)\cdots g(\chi_{n})/\sqrt{q^{n+1}}\cdot %
\ket{b_0,\dots,b_{n}}\otimes\ket{\chi^0,\tilde\chi}$.
\item For every $0\leq i \leq n$, apply the reversible 
$\ket{y}\mapsto\ket{y\cdot c_i^{b_i}}$ to the $\tilde\chi$ register. 
This induces the phase change 
$\ket{\tilde\chi}\mapsto \chi^{-b_i}(c_i)\ket{\tilde\chi} = \ket{\tilde{\chi}}/\chi_i(c_i)$.
\item Finally, apply a general phase flip $(-1)^{n-1}$ to the state. 
\end{enumerate}
By multiplying the phase changes of the above steps, one sees that this algorithm 
establishes the overall evolution
\begin{eqnarray*}
\ket{b_0,\dots,b_{n},\chi^0,\tilde{\chi}}
& \longmapsto & 
\frac{(-1)^{n-1}}{\sqrt{q^{n+1}}}\cdot
\frac{g(\chi_0) \cdots g(\chi_{n})}{\chi_0(c_0) \cdots \chi_{n}(c_{n})}
\ket{b_0,\dots,b_{n},\chi^0,\tilde{\chi}},
\end{eqnarray*}
which describes indeed the roots of Equation~\ref{eq:fermatroots}.
The time and space complexity of the algorithm is $O(n\cdot\poly(\log q))$.
\qed \end{algorithm}

\section{Zeta Functions and Approximate Point Counting}
In this section we look at the relevance of Zeta functions to the
computational task of point counting.  Directly from the
definition $\Zeta_V(T) := \exp(\sum_s N_s T^s/s)$ it is clear that
knowledge about $Z_V$ implies knowledge about the numbers of solutions
$N_s$.  Note for example that we have $\Zeta_V(\delta) = 1 + N_1
\delta + O(\delta^2)$; hence the value of first derivative $\d/\d T$
of $\Zeta_V(T)$ at $T=0$ answers the question ``$N_1=~?$'' 

On the one hand this is good news because counting is a central
problem in computational complexity theory about which we already have
many results.  On the other hand, we know that exact counting very
quickly becomes $\sharpP$-complete (where ``very quickly'' means
determining $N_1$ for moderately complicated $V$).  Because it is
unlikely that quantum computers can solve $\sharpP$-complete
problems \cite{FR}, this makes it at least as unlikely that we will be
able to efficiently determine $\Zeta_V(T)$ exactly with a quantum
algorithm.  Fortunately, the proposal in this article concerns the
design of algorithms that do not try to exactly count the values
$N_s$, but rather only try to \emph{approximate} these quantities.  Hence,
the hardness of $\sharpP$-complete problems does not directly
contradict our spectral approach to Zeta functions.

\subsection{Zeroes of Zeta Functions and Approximate Point Counting.}
By taking the $s$th derivative $\d^s/\d T^s$ at $T=0$ of the logarithm 
of $\Zeta_V(T)$, we see, using Theorem~\ref{thm:weil}, that for the number 
of solutions $N_s$ of $V$ over the finite field extension $\F_{q^s}$ we have 
\begin{eqnarray*}
N_s & = & q^{ds}+1 + \sum_{i=1}^{2d-1}{(-1)^i\sum_{j=1}^{B_i}{\alpha_{i,j}^s}}.
\end{eqnarray*}

For a hypersurface $H$ defined by a homogeneous polynomial 
$f\in\F[X_0,\dots,X_n]$, where most of the roots $\alpha_{i,j}$ are trivial,
this equation becomes especially simple (see Equation~\ref{eq:zetahyper}):
\begin{eqnarray*}
N_s & = & \frac{q^{sn}-1}{q^s-1}
 - (-1)^{n}\sum_j{\alpha_j^s},
\end{eqnarray*}
where the summation goes over the $\approx \deg(f)^n$ non-trivial 
zeroes of $\Zeta_H$, each obeying $\alpha_j := \sqrt{q^{n-1}}\cdot\e^{\i\theta_j}$.
The number of points of an $(n-1)$-dimensional plane in $\P^n(\F_{q^s})$ equals 
the first term $(q^{sn}-1)/(q^s-1)$, hence the $\alpha^s_j$ values express how 
much the number of points of $H$ deviates from those of a straight plane.  
If we assume that we have a unitary transformation $U_H$ whose spectrum 
consists of the $\e^{\i\theta_j}$ phases, then we can express this 
deviation as 
\begin{eqnarray*}
  N_s - \frac{q^{sn}-1}{q^s-1} & = & - (-1)^n \sqrt{q^{s(n-1)}}\cdot\Tr(U_H^s).
\end{eqnarray*}
Hence by estimating the trace of $U_H$ we obtain a non-trivial 
estimation of the number $N_1$ of solutions to the equation $f=0$ over $\F_q$. 

\subsection{Potential Quantum Algorithms for Approximate Counting}
Assume that we can efficiently implement the unitary transformation $U_H$
of the previous subsection.  Here, ``efficiently'' means that the 
time/space complexity of the implementation is $\poly(\log q,\log(\dim(U_H)))
= \poly(\log q,n\log(\deg(f)))$.  Using standard phase estimation 
techniques and $1/\varepsilon$ repetitions, this enables us to estimate 
$-1\leq \Tr(U_H)/\dim(U_H)\leq 1$ with precision $\varepsilon$.
Overall, this gives an estimate $\tilde{N_1}$ with expected error 
\begin{eqnarray*}
\left|{N_1-\tilde{N}_1}\right| &\sim & 
\sqrt{q^{n-1}}\deg(f)^n\cdot\varepsilon,
\end{eqnarray*}
with time/space complexity $\poly(1/\varepsilon,\log q,n\log(\deg(f)))$.  

If one classically (and trivially) samples $1/\varepsilon$ times 
the space $\P^n(\F)$ to estimate the number $N_1$, then the estimated 
error will be $\sim q^n\cdot\varepsilon$, while 
recent results \cite{HI,Schoof} indicate that it might be possible 
to classically count $N_1$ exactly in time $\poly(\log q,\deg(f)^n)$.

This indicates that the conjectured quantum algorithm that uses 
$U_H$ to approximately count $N_1$ would outperform classical
computation in the case where $\deg(f)^n$ is exponentially
big while $\deg(f)$ is smaller than $\sqrt{q}$.
The algorithm that was described in Theorem~\ref{thm:qa} is 
an example of such a case if we fix both $\deg(f)$ and $q$ 
with $\deg(f) < \sqrt{q}$.

\section{Conclusion}
The open problem that remains is obvious: 
For what other hypersurfaces $H$ can we construct quantum
circuits with eigenvalues corresponding to the roots of the
Zeta function $\Zeta_H(T)$?  We consider such circuits efficient
if the space/time complexity is bounded by $\poly(\log q,n\log(\deg(f)))$,
where $f\in\F_q[X_0,\dots,X_n]$ is the homogeneous polynomial that defines $H$.
In addition, if we want to know how useful such quantum circuits are in comparison
with classical algorithms, we also need to know what the classical complexity 
is of approximating roots of Zeta functions.

\subsubsection*{Acknowledgements}
I thank Vinay Deolalikar for explaining to me various aspects of 
algebraic geometry and Zeta functions.

\appendix
\section{Varieties and Algebraic Geometry}
In this section I will give a brief overview of the ingredients from 
commutative algebra and algebraic geometry that are necessary to 
appreciate Weil's Riemann conjectures.  To make matters a bit easier 
the definitions are only done in the context of afine varieties $V$ 
over finite fields $\F$.  For the definition of the Zeta function we 
only deal with the case when the variety is a planar curve $C$.
This restriction allows us to leave out various part of the general
theory. 

\subsection{Ideals and Varieties}
(See \cite{CLO,ReidAG,ReidCA}.)
An \emph{ideal} of the ring of polynomials $\F[X_1,\dots,X_n]$ is a
subset $\mathfrak I \subseteq \F[X_1,\dots,X_n]$ such that: 
\begin{itemize}
\item $0\in \mathfrak I$
\item If $f,g\in \mathfrak I$ then $f+g \in \mathfrak I$
\item If $f\in \mathfrak I$ and $h\in\F[X_1,\dots,X_n]$ then
$fh\in\mathfrak I$.
\end{itemize}  
The ideal \emph{generated} by a finite set of
polynomials $f_1,\dots,f_t$ is the smallest possible ideal $\mathfrak
I\subseteq \F[X_1,\dots,X_n]$ such that $f_1,\dots,f_t\in \mathfrak I$.
This ideal equals
\begin{eqnarray*}
\langle f_1,\dots,f_t\rangle & := & 
\left\{f_1h_1+\dots+f_th_t : h_1,\dots,h_t \in \F[X_1,\dots,X_n]
\right\}.
\end{eqnarray*}
\emph{Hilbert's basis theorem} says that each ideal $\subseteq \F[X_1,\dots,X_n]$
can be finitely generated by a list of polynomials $f_1,\dots,f_t$.

An \emph{affine variety} $\subseteq \A^n(\bar{\F})$ is defined by the
 polynomials $f_1,\dots,f_t$ as the set of roots
\begin{eqnarray*}
\V(f_1,\dots,f_t) & := & \{x\in\A^n(\bar{\F}):f_1(x)=0,\dots,f_t(x)=0\}.
\end{eqnarray*}
For an affine variety $V$ we define the corresponding ideal by 
\begin{eqnarray*}
\I(V) & := & \left\{f\in\F[X_1,\dots,X_n] : f(x)=0 \text{ for all } x\in V\right\}.
\end{eqnarray*}

The crux of the matter here is that for a given variety $V$ there are
many different sets of polynomials $f_1,\dots,f_t$ such that 
$V=\V(f_1,\dots,f_t)$, but there is only one ideal $\I(V)$.  
Hence when studying $V$ it is often more useful to look at $\I(V)$
rather than an arbitrary set of polynomials with $V=\V(f_1,\dots,f_t)$.

We start with a few facts about multivariate polynomials 
over finite fields.  Let $f\in\F[X_1,\dots,X_n]$.  
It is important to realize that ``$f(X)=0$ for all $X\in\A^n(\F)$'' 
does not imply that $f=0$. Take for example $f(X_1) = X_1^2-X_1$ 
for $X_1\in\F_2$. If, however, ``$f(X)=0$ for all $X\in\A^n(\bar{\F})$''
with $\bar{\F}$ algebraically closed, then indeed $f=0$.
Similarly, if $f(X)\neq 0$ for all $X\in\A^n(\bar{\F})$, 
then $f=c \in \F^\times$.
From this it follows that if the variety of a principal 
ideal 
$\langle f \rangle \subseteq \F[X_1,\dots,X_n]$ 
is a subset of $\A^n(\bar{\F})$, then
 $\V(\langle f \rangle)=\A^n(\bar{\F})$ implies $f=0$, and 
 $\V(\langle f \rangle)=\{\}$ implies $\langle f \rangle = \F[X_1,\dots,X_n]$
(because $c\in\F^\times$ generates the whole ring $\F[X_1,\dots,X_n]$).

For any variety $V$ we have $\V(\I(V))=V$.
However, another important issue is the fact that although 
$\langle f_1,\dots,f_t\rangle \subseteq \I(\V(f_1,\dots,f_t))$,
this inclusion is sometimes strict. 
Consider for example
$\langle X^2\rangle \subseteq \I(\V(X^2)) = 
 \langle X \rangle \neq \langle X^2 \rangle$.
Hence for some ideals $\mathfrak I$ we have 
$\I(\V(\mathfrak I))\neq \mathfrak I$.
The following is a necessary and sufficient condition
on $\mathfrak I$ such that $\I(\V(\mathfrak I))=\mathfrak I$.
The \emph{radical ideal} of an ideal $\mathfrak I$
is the ideal 
\begin{eqnarray*} 
\sqrt{\mathfrak I} &:=& \{f : f^m \in \mathfrak I \text{~for some $m\in\Z^+$}\}. 
\end{eqnarray*} 
The \emph{Nullstellensatz} says that for all ideals 
$\I(\V(\mathfrak I))=\sqrt{\mathfrak I}$.  Or, in other words:
$\I(\V(\mathfrak I))=\mathfrak I$ if and only if $\mathfrak I$ is 
\emph{radical.}  A direct consequence of this ``zeroes theorem''
is that $\V(\mathfrak I)=\{\}$ implies $\mathfrak I = \F[X_1,\dots,X_n]$.

A variety $V$ is \emph{irreducible} if $V=V_1\cup V_2$ with $V_1$ and
$V_2$ varieties implies $V=V_1$ or $V=V_2$.  An ideal $\mathfrak
P\subseteq \F[X_1,\dots,X_n]$ is \emph{prime} if for all $fg\in
\mathfrak P$ with $f,g\in\F[X_1,\dots,X_n]$ we have $f\in\mathfrak P$
or $g\in\mathfrak P$.  A variety $V$ is irreducible if and only if the
ideal $\I(V)$ is prime.  Clearly, each prime ideal is radical.

For the algebraically closed field $\bar{\F}$ with varieties
$V\subseteq\A^n(\bar{\F})$ it also holds that for prime ideals
$\mathfrak P$ we have $\I(\V(\mathfrak P))=\mathfrak P$.  Hence we see
that $\I$ and $\V$ establish a bijection between the prime ideals
$\mathfrak P\subseteq \F[X_1,\dots,X_n]$ and the irreducible varieties
$V\subseteq\A^n(\bar{\F})$.  (That this is not true for varieties over
finite fields is readily seen by $\mathfrak P = \langle X^2+X+1
\rangle \subseteq \F_2[X]$ with $\V(\mathfrak P) = \{\} \subseteq
\A^1(\F_2)$ and hence $\I(\V(\mathfrak P))= \F_2[X]\neq \mathfrak
P$. In the case of varieties $\subseteq \A^1(\bar{\F}_2)$ we have
$\V(\mathfrak P) = \F_4\!\setminus\! \F_2$, and hence indeed
$\I(\V(\mathfrak P))=\mathfrak P$.)
For each prime ideal $\mathfrak P$, the quotient ring 
$\F[X_1,\dots,X_n]/\mathfrak P$ is an
\emph{integral domain,} which is a commutative ring with a ``$1$'' and
without zero-divisors.  

\subsection{Addition and Multiplication of Ideals}
One can add and multiply ideals $\mathfrak I,\mathfrak L \subseteq R$
according to $\mathfrak I + \mathfrak L := \{f+g:f\in
\mathfrak I,g\in\mathfrak L\}$ and $\mathfrak I \cdot \mathfrak L := \{\sum_i
f_ig_i : f_i\in\mathfrak I,g_i\in\mathfrak L\}$, where $\sum_i$ is a
finite summation.  Both constructions are again ideals with the
inclusions $\mathfrak{IL}\subseteq \mathfrak I,\mathfrak L\subseteq
\mathfrak I + \mathfrak L$.  The corresponding varieties obey the
rules: $\V(\mathfrak I + \mathfrak L) = \V(\mathfrak I) \cap
\V(\mathfrak L)$ and $\V(\mathfrak I \cdot \mathfrak L) = \V(\mathfrak
I) \cup \V(\mathfrak L)$.  The zero ideal is $\{0\}$, while the unit
ideal is $\langle 1 \rangle = \F[X_1,\dots,X_n]$.

\subsection{Coordinate Ring of a Planar Curve} \label{sec:comalg}
Let $f\in\F[X,Y]$ define a planar curve such that $\langle f \rangle$
is a prime ideal and let $C := \V(\langle f\rangle) \subseteq \A^2(\bar{\F})$ 
be the corresponding irreducible variety.
Let $\F[C] := \F[X,Y]/{\langle f\rangle}$ be the commutative \emph{coordinate 
ring} of polynomials on the curve $C$. 
Because $\langle f\rangle$ is prime, $\F[C]$ is an \emph{integral domain,}
which means that $gh=0$ implies $g=0$ or $h=0$ in $\F[C]$.
In fact, the commutative ring $\F[C]$ is a \emph{Dedekind ring} such that 
all ideals $\mathfrak I\subseteq \F[C]$ will have a unique
factorization in terms of prime ideals $\mathfrak P \subseteq \F[C]$.  

How do the ideals $\mathfrak I$ of $\F[C]$ look like?  Let $g \in \mathfrak I$
with a point $(a,b)\in\A^2(\bar{\F})$ on the curve $C$ such that
$g(a,b)=0$. Define the \emph{Frobenius automorphism}
$\phi:\bar{\F}\rightarrow \bar{\F}$ by $\phi:z\mapsto z^q$, where $q$
is the size $|\F|$ of the base field.  It is straightforward to see
that $\phi$ is indeed an automorphism with 
$\phi(x+y)=\phi(x)+\phi(y)$ and $\phi(xy)=\phi(x)\phi(y)$ for all
$x,y\in\bar{\F}$.  The $r$th power $\phi^r$ is defined by $\phi^r(z)
:= z^{(q^r)}$, such that $\phi^r$ acts as the identity on $\F_{q^r}$.
Let $\phi(a,b)$ denote the point $(\phi(a),\phi(b))$.  Because
$g\in\F[X,Y]$ we see that $g(a,b)=0$ implies $g(\phi^r(a,b))=0$ for
$r=\Z^+$.  Let $d$ be the \emph{degree} of the point $(a,b)$,
which means that $d$ is the smallest integer such that
$(a,b)\in\A^2(\F_{q^d})$.  Then $\phi^d(a,b)=(a,b)$ and the \emph{orbit}
$O_{(a,b)} := \{(a,b), \phi(a,b), \phi^2(a,b),\dots,\phi^{d-1}(a,b)\}$ 
consists of $d$ different points in $\A^2(\bar{\F})$.  Because $f\in\F[X,Y]$, all
points $\phi^r(a,b)$ will also lie on the curve $C$.  As a result, $g$
will also be zero on $C$ for all $d$ points $(a,b),
\phi(a,b),\dots,\phi^{d-1}(a,b)$.  This result extends to the ideals
$\mathfrak I\subseteq\F[C]$: for each point $(a,b)\in C$ with
$(a,b)\in\V(\mathfrak I)$, the whole orbit lies in the variety:
$\{(a,b),\phi(a,b),\dots,\phi^{d-1}(a,b)\}\subseteq \V(\mathfrak I)$.

Conversely, let $S\subseteq C$ be a set of points that is closed under
the Frobenius automorphishm (that is: if $(a,b)\in S$, then
$\phi(a,b)\in S$).  It is straightforward to check that the set of
polynomials $\{g \in \F[C] : g(a,b)=0 \text{ for all }(a,b)\in S\}$ is
an ideal in $\F[C]$.

The prime ideals of $\F[C]$ are in one-on-one correspondance with 
unique orbits in $C$.  For each orbit $O_{(a,b)}\subset C$
the ideal $\I(O_{(a,b)})$ is prime, and for each prime ideal 
$\mathfrak P\subset \F[C]$ the variety $\V(\mathfrak P)\subset C$
is an orbit $O_{(a,b)}$.

\subsection{Norms in Coordinate Rings}
The polynomial $f\in\F[X,Y]$ is \emph{absolutely irreducible} 
if and only if $f$ is irreducible in $\bar{\F}[X,Y]$.  
A point $(a,b)$ of the corresponding curve $C$ is \emph{singular} 
if and only if $\d f/\d X = \d f/\d Y = 0$ at $(a,b)$.
The curve $C$ is \emph{nonsingular} if and only if it has no 
singular points.

If $f$ is absolutely irreducible and the corresponding curve $C$
is nonsingular, then we can define a norm on the ideals of 
$\F[X,Y]/(f)$ as follows.
Let $\mathfrak I$ be a non-zero ideal, then its norm in the
coordinate ring $\F[C]$ is defined by 
$\|\mathfrak I\| := |\F[C]/\mathfrak I|$, 
which will always be a finite number.  
Recalling the definition of multiplication of ideals, one can
show that $\|\mathfrak I\cdot\mathfrak L\| = 
\|\mathfrak I\|\cdot\|\mathfrak L\|$.
For each prime ideal $\mathfrak P$, we have that 
$\|\mathfrak P\| = q^d$ where $d$ is the size of the
orbit corresponding to $\mathfrak P$.

\section{Why It Is Called a Zeta Function}
Here I will give a brief explanation of the connection between
Riemann's zeta function and the Zeta function of finite field
equations.  

\subsection{Riemann's Zeta Function}
For $z\in\C$ with $\Re(z)>1$, Riemann's zeta function is defined by
$\zeta_R(z) := \sum_{n=1}^{\infty}{n^{-z}}$.  For the other $z\in \C$
with $\Re(z)\leq 1$, the zeta function $\zeta(z)$ is determined by the
analytical continuation of the $\Re(z)> 1$ part of the function.  This
continuation gives explicitly $\zeta(z) =
(\sum_{n=1}^{\infty}{(-1)^{n-1}/n^z})/(1-2^{1-z})$ for $\Re(z)>0$ and
$\zeta(1-z) = 2(2\pi)^{-z}\cos(z\pi/2)\Gamma(z)\zeta(z)$ for
$\Re(z)<0$.  (Compare this with the function
$f(z)=\sum_{n=1}^{\infty}{z^{-n}}$, which is properly defined only for
$|z|>1$, but which can be continued to the whole $z\in\C$ plane by the
function $1/(z-1)$.)

The Riemann zeta function diverges to $+\infty$ as $s\rightarrow 1$
and $\zeta(z)=0$ for $z=-2,-4,-6,\dots$ (the \emph{trivial zeroes}).
The remaining roots of $\zeta$ lie in the strip $0<\Re(z)<1$.  The
\emph{Riemann conjecture} states that all these non-trivial zeroes $z$
are in fact on the line $\Re(z)=\frac{1}{2}$.  Some well-known
function values are: $\zeta(-2)=0$, $\zeta(-1)=-\frac{1}{12}$,
$\zeta(0)=-\frac{1}{2}$, $\zeta(1)=+\infty$, $\zeta(2)={\pi^2}/{6}$,
$\zeta(3)=1.202056903\dots$, and the first non-trivial root
$\zeta(\frac{1}{2}\pm \i \cdot 14.1347\dots)=0$.
 
The connection between prime numbers and the zeta 
function is given by the important \emph{Euler product} 
\begin{eqnarray*}
\zeta(z) & = & \sum_{n=1}^{\infty}{\frac{1}{n^z}}\\
&=& \prod_{p=2,3,5,\dots}{\left({1+\frac{1}{p^z}+\frac{1}{p^{2z}}+\dots}\right)} \\
& = & \prod_{p=2,3,5,\dots}{\frac{1}{1-p^{-z}}}
\end{eqnarray*} 
where the product is over all primes $p$ in $\Z$.
For example, this shows how the value $\zeta(1)=+\infty$ implies 
the fact that there is an infinite number of primes in $\Z$.
The Euler product relies crucially on the fact that each number $n=1,2,3,\dots$
has a \emph{unique factorization} in terms of the primes $p=2,3,5,\dots$.

The ideals of $\Z$ are the subsets $\mathfrak N = n\Z$ with the norm
$\|{\mathfrak N}\|:=|\Z/n\Z|= n$.  The zeta function is hence defined
by $\zeta(\Z,z) := \sum_{\mathfrak N}{\|\mathfrak N\|^{-z}} =
\prod_{\mathfrak P}{1/(1-\|\mathfrak{P}\|^{-z})}$, which relies on the
unique factorization of each ideal $\mathfrak N$ into a sequence of
prime ideals $\mathfrak P$:
\begin{eqnarray*}
\sum_{\mathfrak N\subseteq \Z}{\frac{1}{\|\mathfrak N\|^z}} 
& = & 
\prod_{\mathfrak P \subseteq \Z}{\frac{1}{1-\|\mathfrak P\|^{-z}}}.
\end{eqnarray*}

\subsection{Dedekind Zeta Functions}
For number fields $K=\Q(\theta)$ other than $\Q$ we can extend the
notion of a zeta function to that of \emph{Dedekind
zeta functions.}  We thus get a zeta function for the ring $\Z_K$
(like the Gaussian integers $\Z[\i]$) that are defined similar to the
previous equation, except for that the summation (product) now ranges
over the (prime) ideals $\mathfrak N,\mathfrak P\subseteq \Z_K$.
\begin{example}[Dedekind Zeta Function of Gaussian Integers]\label{ex:gaussint}
Consider the elements in the ring $\Z[\i] := \{a+b\i : a,b\in \Z\}$,
which is a \emph{principal ideal domain} with units $\pm 1$ and $\pm
\i$.  Each ideal $(a+b\i)$ can be viewed as a lattice in $\Z[\i]$ in
that is spanned by the vectors $a+b\i$ and $-b+a\i$, which shows that
that we have the norm $\|(a+b\i)\| := |\Z[\i]/(a+b\i)|=a^2+b^2$ on the
ideals of $\Z[\i]$.

The prime ideals $\mathfrak P\subset \Z[\i]$ are described as 
follows (see Section~9.7 in \cite{IR}):
\begin{itemize}
\item $\mathfrak P = (1+\i)$, with obviously $\|\mathfrak P\|=2$.
\item $\mathfrak P = (p)$ with $p=3\bmod{4}$ a prime in $\Z$, 
such that $\|\mathfrak P\|=p^2$.
\item For each prime $p=1\bmod{4}$ in $\Z$ there are two 
different prime ideals $(a+b\i)$ and $(a-b\i)$ 
(determined by $a^2+b^2=p$), such that for both ideals 
$\|\mathfrak P\| = p$.
\end{itemize} 
Hence the Dedekind zeta function of $\Z[\i]$ equals
\begin{eqnarray*}
\zeta(\Z[\i],z) & := & 
\prod_{\mathfrak P \subset \Z[\i]}{\frac{1}{1-\|\mathfrak P \|^{-z}}}\\
& = & 
\frac{1}{1-2^{-z}}
\prod_{p=3,7,11,\dots}{\frac{1}{1-p^{-2z}}}
\prod_{p=5,13,17,\dots}{\frac{1}{(1-p^{-z})^2}}
\end{eqnarray*}
\end{example}

The conjecture that the non-trivial zeroes of zeta functions $\zeta_K$ 
of all such $K$ lie on the $\Re(z)=\smfrac{1}{2}$ line is known
as the \emph{extended Riemann hypothesis.}

\subsection{Zeta Functions of Algebraic Varieties}
Let $f\in\F[X,Y]$ be an absolutely irreducible function 
and let $C\subseteq \A^2(\bar{\F})$ be its nonsingular curve.
In Section~\ref{sec:comalg} we mentioned that this implies that 
the ring $\F[C] := \F[X,Y]/\langle f\rangle$ is a Dedekind domain with
finite quotients.  This means that the nontrivial ideals 
$\mathfrak I\subseteq \F[C]$ have a unique factorization
$\mathfrak I = {\mathfrak P}_1\cdots {\mathfrak P}_s$ 
into the prime ideals ${\mathfrak P}_i$ of $\F[C]$.
Furthermore, for each nonzero ideal the norm $\|\mathfrak I\| 
:= |\F[C]/\mathfrak I|$ will be a finite integer.

We can now define the zeta function of $\F[C]$ by
\begin{eqnarray*}
\zeta(\F[C],z) & := & \sum_{\mathfrak I}{\frac{1}{\|\mathfrak I\|^z}},
\end{eqnarray*} 
where we sum over all non-zero ideals of $\F[C]$.
Because of the unique factorization property of $\F[C]$, 
we can rewrite the zeta function according to the Euler product
\begin{eqnarray*}
\zeta(\F[C],z) & = & \prod_{\mathfrak P}{\frac{1}{1-\|\mathfrak P\|^{-z}}},
\end{eqnarray*}
with the product over all prime ideals $\mathfrak P$ of $\F[C]$. 

As mentioned in Section~\ref{sec:comalg}, 
each prime ideal $\mathfrak P$ corresponds to an orbit
$O_{(a,b)}=\{{(a,b),\phi(a,b),\dots,\phi^{d-1}(a,b)}\}$ of $d$ 
points in $C$, where $d$ is the degree of the point $(a,b)$.  
The norm of such a prime ideal equals 
$\|\mathfrak P\|= q^d$.  Let $b_d$ be the number of ideals that
have norm $q^d$ and define $\Zeta(\F[C],q^{-z}) := \zeta(\F[C],z)$ in
combination with the substitution $T\leftarrow q^{-z}$, then
\begin{eqnarray*}
\Zeta(\F[C],T) & = & \prod_{d=1}^{\infty}{(1-T^{d})^{-b_d}}.
\end{eqnarray*}
Taking the natural logarithm of this infinite product 
gives the following summation 
\begin{eqnarray*}
\log(\Zeta(\F[C],T)) & = & -\sum_{d=1}^{\infty}{b_d \log(1-T^d)}\\
& = & 
\sum_{d,j=1}^{\infty}{b_d \frac{T^{dj}}{j}}.
\end{eqnarray*}
By looking at fixed powers $T^s$, we can rewrite this summation as
\begin{eqnarray*}
\log(\Zeta(\F[C],T)) & = & 
\sum_{s=1}^{\infty}{\sum_{d|s}{d b_d \frac{T^{s}}{s}}}.
\end{eqnarray*}
Now, to count the number of points $N_s := C\cap\A^2(\F_{q^s})$, 
we add all orbits that have coefficients in $\F_{q^s}$.
Because a degree $d$ orbit lies in $\A^2(\F_{q^s})$ if and only 
if $d|s$ (as $F_{q^d}\subseteq \F_{q^s}$ if and only if $d|s$),
we see that $N_s =\sum_{d|s}{d b_d}$.  
Going back to the Zeta function, we thus have 
\begin{eqnarray} \label{eq:appzeta}
\Zeta(\F[C],T) & = & 
\exp\left({\sum_{s=1}^{\infty}{N_s \frac{T^s}{s}}}\right).
\end{eqnarray}
for $C\subseteq \A^2(\bar{\F})$ a nonsingular curve defined by
a completely irreducible polynomial $f\in\F[X,Y]$. 
Note however that---with a leap of faith---Equation~\ref{eq:appzeta} 
can also be used to define a Zeta function $\Zeta(\F[V],T)$ 
for general affine varieties $V\subseteq \A^n(\bar{\F})$ 
or projective varieties $V\subseteq \P^n(\bar{\F})$. 

\subsection{Connection Between Local and Global Zeta Functions}
For each prime field $\Z/p\Z$ we say that $\zeta(\Z/p\Z,z) =
1/(1-p^{-z})$ for the following reason.  Consider $\Z/p\Z \cong
\F_p[X]/(X)$ where $X=0$ has one solution ($N_s=1$) for all fields
$\F_{p^s}$. Using the definition of the Zeta function for varieties,
we find that $\Zeta(\F_p,T) = 1/(1-T)$ for all $p$, and hence indeed
$\zeta(\F_p,z) = 1/(1-p^{-z})$.  Observe now that we can recover the
Riemann zeta function (defined for $\Z$) by the product
\begin{eqnarray*}
\zeta(\Z,z) & = & \prod_{p=2,3,5,\dots}{\zeta(\Z/p\Z,z)},
\end{eqnarray*}
over all primes $p$.  We say that 
the \emph{global Riemann zeta function} over $\Z$
is composed by the \emph{local zeta functions} over $\Z/p\Z$.
The next example for the Gaussian integers
$\Z[\i]$ indicates that this is no coincidence.

\begin{example}[Local and Global Zeta Functions of Gaussian Integers]
Because $\Z[\i]\cong \Z[X]/(X^2+1)$, we look at the 
zeta functions of the equation $X^2=-1$ for the various 
base fields $\F_p$:
\begin{itemize}
\item $p=2$: The equation $X^2=-1$ has one solution, hence 
$\zeta(\F_2[X]/(X^2+1),z) = 1/(1-2^{-z})$.  
\item  $p\neq 2$ and $p=1\bmod{4}$: The equation $X^2=-1$
has two solutions in $\F_{p^s}$, hence 
$\zeta(\F_p[X]/(X^2+1),z) = 1/(1-p^{-z})^2$.
\item $p\neq 2$ and $p=3\bmod{4}$: The equation $X^2=-1$
in $\F_{p^s}$ has no solution if $s$ is odd, and 
it has two solutions if $s$ is even. Hence 
(using $N_s=1+(-1)^s$) we have  $\zeta(\F_p[X]/(X^2+1),z) = 1/(1-p^{-2z})$.
\end{itemize}
Altogether, this gives the product 
\begin{eqnarray*}
\prod_{p=2,3,5,\dots}{\zeta(\F_p[X]/(X^2+1),z)} & = & 
\frac{1}{1-2^{-z}}
\prod_{p=5,13,\dots}{\frac{1}{(1-p^{-z})^2}}
\prod_{p=3,7,\dots}{\frac{1}{1-p^{-2z}}},
\end{eqnarray*}
which corresponds exactly with the zeta function of 
$\Z[\i]\cong \Z[X]/(X^2+1)$ in Example~\ref{ex:gaussint}. 
\end{example}

\begin{thebibliography}{99}
\bibitem{BK}
Michael Berry and Jon Keating, 
``$H=xp$ and the Riemann zeros'', 
\emph{Supersymmetry and Trace Formulae: Chaos and Disorder,}
pp.~355--367, Plenum Press (1999)

\bibitem{revisited}
Richard Cleve, Artur Ekert, Chiara Macchiavello, and Michele Mosca,
``Quantum algorithms revisited'',
\emph{Proceedings of the Royal Society of London A,}
Volume~454, pages~339--354 (1998);
arXiv:quant-ph/9708016

\bibitem{CLO}
David Cox, John Little and Donal O'Shea, 
\emph{Ideals, Varieties, and Algorithms: An 
Introduction to Computational Algebraic Geometry and
Commutative Algebra,} second Edition, 
Undergraduate texts in mathematics, 
Springer-Verlag (1997)

\bibitem{vD}
Wim van Dam, 
``Quantum Computing Discrete Logarithms with the Help of a Preprocessed State'',
arXiv:quant-ph/0311134 (2003)

\bibitem{vDS1}
Wim van Dam and Gadiel Seroussi, 
``Efficient Quantum Algorithms for Estimating Gauss Sums'', 
arXiv:quant-ph/0207131 (2002) 

\bibitem{vDS2}
Wim van Dam and Gadiel Seroussi, 
``Quantum Algorithms for Estimating Gauss Sums and Calculating Discrete Logarithms'', 
in submission (2003).

\bibitem{Dieu} Jean Dieudonn\'e,
``On the History of the Weil Conjectures'', 
\emph{The Mathematical Intelligencer}, Volume~10, Spinger (1975); 
also in Eberhard Freitag and Rinhardt Kiehl, 
\emph{Etale Cohomology and the Weil Conjectures,}
Ergenisseder Mathematik Und Ihre Grenzgebiete, Volume~13, Springer Verlag (1988) 

\bibitem{FR}
Lance Fortnow and John Rogers, 
``Complexity limitations on quantum computation'', \emph{Proceedings of the 13th
IEEE Conference on Computational Complexity,} pp.~202--209 (1998);
arXiv:cc.CC/9811023 

\bibitem{Harts}
Robin Hartshorne, 
\emph{Algebraic Geometry,} % Appendix C
Graduate Texts in Mathematics, Volume~52, Springer-Verlag (1977) 

\bibitem{HI}
Ming-Deh Huang and Doug Ierardi, 
``Counting Points on Curves over Finite Fields'', 
Journal of Symbolic Computation, Volume~25, pp.~1--21 (1998)

\bibitem{IR} 
Kenneth Ireland and Michael Rosen,
\emph{A Classical Introduction to Modern Number Theory,}
Second Edition, Springer, Graduate Texts in Mathematics 84 (1990) 

\bibitem{KZ1}
Nicholas M. Katz and Peter Sarnak, 
``Zeroes of Zeta Functions and Symmetry'', 
\emph{Bulletin of the American Mathematical Society,}
Volume~36, Number~1, pp.~1--26 (1999)

\bibitem{KZ2}
Nicholas M. Katz and Peter Sarnak, 
\emph{Random matrices, Frobenius eigenvalues, and monodromy,}
American Mathematical Society Colloquium Publications, 
Volume 45 (1999)

\bibitem{KLR}
Jon Keating, Noah Linden, and Zeev Rudnick, 
``Random matrix theory, the exceptional Lie groups and $L$-functions'',
\emph{Journal of Physics A: Mathematical and General,}
Volume~36, pp.~2944--2944 (2003)

\bibitem{Kitaev}
Alexei Yu. Kitaev, Alexander Shen, and Mikhail N. Vyalyi, 
\emph{Classical and Quantum Computation,}
Graduate Studies in Mathemetics, Volume 47, 
American Mathematical Society (2002)

\bibitem{Koblitz}
Neal Koblitz, 
``The Number of Points on Certain Families of Hypersurfaces over Finite Fields'',
\emph{Compositio Mathematica,} Volume.~48, No.~1, pp.~3--23 (1983)

\bibitem{Lorenzini}
Dino Lorenzini, 
\emph{An Invitation to Arithmetic Geometry,}
Graduate Studies in Mathematics, Volume~9, 
American Mathematical Society (1996)

\bibitem{NC}
Michael A.\ Nielsen and Isaac L.\ Chuang,
\emph{Quantum Computation and Quantum Information,} 
Cambridge University Press (2000)

\bibitem{Odlyzko}
Andrew M. Odlyzko, 
``The $10^{22}$-nd zero of the Riemann zeta function'',
\emph{Contemporary Mathematics,} Volume~290, pp.~139--144 
American Mathematical Society (2001)

\bibitem{ReidAG}
Miles Reid, 
\emph{Undergraduate Algebraic Geometry,}
London Mathematical Society Student Texts, 
Volume~12, Cambridge University Press (1988)

\bibitem{ReidCA}
Miles Reid, 
\emph{Undergraduate Commutative Algebra,}
London Mathematical Society Student Texts, 
Volume~29, Cambridge University Press (1995)

\bibitem{Rosen}
Michael Rosen, 
\emph{Number Theory in Function Fields,}
Graduate Texts in Mathematics, Volume 210, Springer-Verlag (2002) 
%Appendix: A Proof of the Function Field Riemann Hypothesis, pp.~329--339

\bibitem{Shor} 
Peter W.\ Shor, 
``Algorithms for Quantum Computation: Discrete  Logarithms and Factoring'', 
\emph{SIAM Journal on Computing,} Volume~26:5,  pp.~1484--1509 (1997);
arXiv:quant-ph/9508027

\bibitem{Schoof} 
Ren\'e Schoof, 
``Elliptic curves over finite fields and the computation of square roots $\mod{p}$'', 
Mathematics of Computation, Volume~44, No.~170, pp.~483--494 (1985)

\bibitem{Thomas}
Alan David Thomas, \emph{Zeta-functions: An introduction to algebraic geometry,}
Research Notes in Mathematics, Volume 12, Pitman Publishing (1977)

\bibitem{Watrous}
John Watrous, ``Quantum algorithms for solvable groups'', 
\emph{Proceedings of the 33rd ACM Symposium on Theory of Computing,} 
pp.~60--67 (2001) 

\bibitem{Weil}
Andr\'e Weil, ``Numbers of solutions of equations in finite fields'',
\emph{Bulletin of the American Mathematical Society,} 
Volume~55, pp.~497--508 (1949)

\bibitem{Zeta}
``Zeta functions'', \emph{Encyclopedic Dictionary of Mathematics,}
edited by S.\ Iyanaga and Y.\ Kawada, Volume~2, pp.~1372--1393, 
Mathematical Society of Japan, MIT Press (1977)
\end{thebibliography}
\end{document}